This is an **excerpt from the book** “World of Movable Objects” which is available together with the whole project of the accompanying application (all codes are in C#) at www.sourceforge.net in the project **MoveableGraphics** (names of projects are case sensitive there!).

# World of Movable Objects

## *Preface*

Suppose that you are sitting at the writing table. In front of you there are books, text-books, pens, pencils, pieces of paper with some short notes, and a lot of other things. You need to do something, you have to organize your working place, and for this you will start moving the things around the table. There are no restrictions on moving all these things and there are no regulations on how they must be placed. You can put some items side by side or atop each other, you can put several books in an accurate pile, or you can move the bunch of small items to one side in a single movement and forget about them, because you do not need them at the moment. You do not need any instructions for the process of organizing your working place; you simply start putting the things in such an order, which is the best for you at this particular moment. Later, if something does not satisfy you or if you do not need some items in their places, you will move some of the items around and rearrange everything in whatever order you need without even paying attention to this process. You make your working place comfortable for your work at each moment and according to your wish.

The majority of those who are going to read this text nearly forgot everything about paper, books, hand writing… A personal computer became the only instrument and the working place for millions of people. Screens of our computers are occupied with different programs; the area of each program is filled with many different objects. Whenever you need to do something, you start the needed program and in each of them you know exactly, what you want to do and what you can do. Did it ever come to your attention that in any program you know exactly the whole set of possible movements and actions? Have you ever understood that the set of allowed actions is extremely small and you try to do your work within a strictly limited area of allowed steps? Those limits are the same in all the programs; that is why there are no questions about the fairness of such situation. You can press a button; you can select a line or several lines in a list; you can do several other typical things, but you never go out of the standard actions which you do in any other program. You and everyone else know exactly what they are allowed to do. Other things are neither tried, nor discussed. They simply do not exist.

Now try to forget for a minute that for many years you were taught what you could do. Let us say that you know, how to use the mouse (press – move – release), but there are no restrictions on what you are allowed to do with a mouse. Switch off the rules of your behaviour “in programs” according to which all your work was going for years.

Try the new set of rules:

- You can press ANY object and move it to any new location.
- You can press the border of any object and change its size in the same easy way; there are no limitations on such resizing (except some obvious and natural).
- A lot of objects you can reconfigure in the same simple way by moving one side or another.

It must be obvious to you that the reaction of a program on your clicking a button does not depend on the screen position of this button, so if you change the size or location of a button or a list it is not going to change any code that is linked with clicking a button, selecting a line in the list or making a choice between several positions. All the programs are still going to work according to their purposes. Buttons and lists represent a tiny part of objects that occupy the screens. Now make one more step and imagine the programs, in which ALL the objects are under your control in exactly the same way. You can do whatever you want with the objects, while the programs continue to work according to their purposes.

What would you say about such a world? Do not be in a hurry with your answer. You never worked with such programs; you would better try before giving an answer. This book is about the screen world of movable and resizable objects. Those objects can be of very different shapes, behaviour, and origin; that is why there are more than 80 examples in the accompanying program. Some of those examples are simple; others are in reality complex and big enough applications by themselves. And there is <u>not a single unmovable object in all of them</u>.

The world of movable objects – the world of *user-driven applications*.

# Contents

# Introduction

Some time ago I wrote two articles to describe the main results of my work throughout the last four years. The first of them appeared in the late fall of 2009 and was called "*On the theory of moveable objects*" [5]. The second article appeared in spring of 2010 and was called "*User-driven applications*" [6]. These papers came in a row with my previous articles [7 - 11] and summarized the results of years of work. Both papers [5, 6] were accompanied by their own demonstration programs and the full codes for both projects. I tried to make those articles as short as possible because they were also published on the site www.codeproject.com, where the average and expected publications are usually short. Even with all my efforts, each of those two articles was around 30 pages. It was the limit below which I could not go; even such volume required the removal of many important explanations. As a result, both texts were left without some needed pieces, though they gave the main idea of my work. To reduce the volume of those articles, I often included into the texts some remarks on the code, but without putting the code itself next to the explanation. The codes can be looked through separately because all of them are available, but this process requires much more efforts from the readers. I got some complaints about it, but I knew these things before publishing the articles, so I had only to agree with those remarks.

The two articles [5, 6] are strongly related and represent the halves of one general problem and its solution. The first article was about "How to do…"; the second was about "What would happen if the previous results were used". The division in two parts was done only due to the technical (publishing) problems; in reality those two things must be always regarded as a single theory. So, this is the main idea of this book: to introduce the whole theory without any divisions and to do it in a more detailed form.

The book is divided into two big parts which correspond to those two articles, but there is some problem in such presentation. The book has a linear structure with the chapters going one after another. Any next chapter can use the explanations from the previous part of the book and introduce something new. At the same time, very few real algorithms have a linear structure; the structure of a tree is much more common in programming than anything else. Each new piece opens the way for a new branch or even several branches of ideas. It would be nice to have the book with the same tree structure, but I do not know how to write such a book, so I continue to write this one in a traditional linear way.

The book is accompanied by a program with a lot of examples. Even from the beginning, the examples are designed according to all the ideas of the user-driven applications, but the explanation of these ideas and the discussion of how they were born and why the applications must be designed in such a way appear much later, in the second half of the book. It cannot be done in another way: before talking about the ideas of user-driven applications I have to demonstrate the design of all the elements, without which such applications simply cannot exist.

I would suggest one thing. If you are not familiar with those two articles [5, 6], look through them first and at the programs that come with them. (Applications and articles are available at www.sourceforge.net.) Look at them at least in a quick way; this will give you some understanding of the whole area of discussion. But be aware that even if later you see the similar examples from the articles in the book, the later versions from the book usually have some significant additions.

This book is about two things:

- The design of movable / resizable objects.
- The development of user-driven applications.

The two theories can be looked at as the independent things because:

- The design of movable objects is not influenced in any way by the afterthoughts of where these objects are going to be used.
- The design of user-driven applications is independent of the real algorithm for constructing movable objects.

At the same time there is a very strong relation between the two things, as the user-driven applications can be designed only on the basis of movable / resizable objects and all the extraordinary features of such applications are the results of their construction exclusively and entirely of movable objects. The movable / resizable objects can be used by themselves in the existing applications of the standard type, but only invention of an algorithm turning an arbitrary object into movable / resizable allowed to design an absolutely new type of programs – user-driven applications.

Often enough, when I tell people that I have thought out an algorithm of making movable any screen object, a lot of people are a bit (or strongly) surprised: "What are you talking about? Is there anything new in it? We have seen objects moving around the screen for years and years?" Certainly, they saw; as the demonstration of moving objects has the history of several decades. Only all those objects moved according to the scenario written by their developers. Users could do nothing about that moving except watching. The thing I was working on for years and which I am going to describe here is absolutely different: it is the moving of screen objects by the users of the programs. This article is about the development of screen objects, movements of which are not predicted by the designer of an application. It is not about a film developed on a predetermined scenario. It is about an absolutely new type of programs – *user-driven applications*. In these programs all objects, from the simplest to the most complicated, consisting of many independent or related parts, are designed to be moved and resized only by USERS. The objects are designed with these special features, and then the whole control of WHAT, WHEN, and HOW is going to appear on the screen is given to the users.

I would like to mention beforehand that the overall behaviour of user-driven applications is so different from whatever you experienced before that you can feel a shock or at least a great amusement at the first try. From my point of view, such a reaction from the people who are introduced to the user-driven applications for the first time supposed to be absolutely natural. A lot of people had the same feeling of a shock when, after years of work under DOS, they tried the Windows system for the first time. By the way, the only visual difference of the Windows system from the familiar DOS was the existence of several movable / resizable windows (in comparison with a single one and unmovable) and the possibility of moving icons across and along the screen. That was all! And even that was a shock. From the users' point of view, the step from the currently used programs to the user-driven applications is much bigger than that old step from DOS to Windows. In user-driven applications EVERYTHING is movable and resizable, and this is done not according to some predefined scenario, but by users themselves.

Movability of elements is not some kind of an extra feature that is simply added to well known objects in order to improve their behaviour. Technically (from the programming point of view), it is adding a new feature, but it turned out to be not simply a small addition to the row of other features. The movability of objects which are well known for years changes the way of using these objects absolutely. The most remarkable result of this change is that the applications have to be redesigned, while movability of all their parts must be taken into consideration. Movable and unmovable elements cannot coexist on the screen in any way; they immediately begin to conflict and demand the transformation of any unmovable objects into movable. Movability of elements changes the whole system of relations between the objects of applications. Movability of objects is the main, the fundamental feature of the new design. .

Throughout the whole history of programming we have a basic rule which was never doubted since the beginning and up till now and, as a result, turned into an axiom: <u>any application is used according to the developer's understanding of the task and the scenario that was coded at the stage of the design</u>. After reading this, you will definitely announce a statement: "Certainly. How else can it be?" Well, for centuries there was a general view, which eventually turned into an axiom, that the Sun was going around the Earth. There were no doubts about it. Yet, it turned out to be not correct.

40 years ago the majority of programs were aimed at solving some scientific or engineering tasks and the overwhelming majority of those programs were written in FORTRAN. I think that not too many readers of this book can explain or even remember the origin of this name; I would remind that it stands for **FOR**mula **TRAN**slation. The main purpose of the language was to translate the equations and algorithms into the intermediate notation, which was, in its turn, translated into the inner machine codes. The main goal of the language was to deal with formulas! Computers were big calculators and nothing else. It was not strange at all that whatever commands there were for visualizing the data and results, those commands were intermixed with other commands for calculations. At that time nobody asked the questions about such development of programs. The author of a program tried to write the consecution of instructions to turn formulas into the final results; it would be nice to see some intermediate results, if the calculations were long and complicated. They were often very long and complicated, so few extra operations for showing out intermediate values were incorporated into the block of calculations just at the places, where those values were obtained.

Years later much better visualization was achieved both with the hardware improvement and the design of new languages. With this there came understanding that calculations and visualization had to be separated. They were separated from the point of programming, but the same person – developer – is still responsible for everything. A developer knows all the insides of the calculations and he decides what, when, and how to show. Eventually this developers' dictate came into conflict with the wide variety of users' requests for visualization; the adaptive interface was thought out to solve the problem. Many forms of adaptive interface were proposed (the dynamic layout is only one of its popular branches), but all those numerous solutions are only softening the problems but not solving them. The main defect of the adaptive interface is in its own base: the designer puts into the code the list of available choices for each and all situations he can think about. Users have no chances to step out of the designer's view and understanding of any situation. It is the dead end of evolution programs under those ideas which were proposed around 25 years ago.

With the movability of all the parts from the tiny elements to the most complex objects, there is another way of application design, when a program continues to be absolutely correct from the point of fulfilling its main purpose (whatsoever this purpose is), but at the same time does not work according to the predefined scenario and does not even need such a scenario. To do such a thing, an application has to be developed not as a program in which whatever can be done with it has to be thought out by the developer beforehand and hard coded; instead an application is turned into an instrument of solving problems in particular area. An instrument has no fixed list of things that can be done with it, but only an idea of how it can be used; then an instrument is developed according to this idea. A user of an instrument has full control of it; only the user decides when, how, and for what purpose it must be used. Exactly the same thing happens with programs that are turned into instruments.

I call the programs, based on movable / resizable objects, ***user-driven applications***. When you get a car, you get an instrument of transportation. Its manual contains some suggestions on maintenance, but there is no fixed list of destinations for this car. You are the driver, you decide about the place to go and the way to go. That is the meaning of the term *user-driven application*: you run the program and make all the decisions about its use; a designer only provides you with an instrument which allows you to drive.

The first half of this book is about the design of movable objects. I have already mentioned the first misunderstanding of the importance of this task, based on not realizing the difference between the moving according to the predefined scenario (it is simply an animation) and the moving of objects according to the user's wish. But when I explain the obvious difference between these two things, I often hear another statement. "Everyone can move and resize the windows at any moment and in any way he wants, so what is the novelty of your approach?" The answer is simple, but a bit longer than on the first question.

Rectangular windows are the basic screen elements of the Windows operating system.* You can easily move all these windows, resize them, overlap them or put them side by side. At any moment you can reorganize the whole screen view to whatever you really need. It was not this way at the beginning of the computer era; it became the law after Windows conquered the world. This is **axiom 1** in modern day programming design: *On the upper level, all objects are movable and resizable*. To make these features obvious and easy to use, windows have title bars by which they can be moved, and borders by which they can be resized. Being movable and resizable are standard features of all the windows, and only for special purposes these features can be eliminated.

Usually the goal of switching on the computer is not to move some rectangular windows around the screen, but to do something special in applications which are represented by those windows. You start an application you need, you step onto the inner level, and then everything changes. Here, inside the applications, you are doing the real work you are interested in, and at the same time you are stripped of all the flexibility of the upper level – you can only do what the designer of the program allows you to do. The design can be excellent or horrible, it can influence the effectiveness of your work in different ways, but still it is awkward that users are absolutely deprived of any control of the situation. Have you ever questioned the cause of this abrupt change? If you have, then you belong to the tiny percentage of those who did. And I would guess that the answer was: "Just because. These are the rules."

Unfortunately, these ARE the rules, but rules are always based on something. The huge difference between the levels is that on the upper level there is only one type of objects – windows, while on the inner level there are two different types: controls, inheriting a lot from windows, and graphical objects that have no inheritance from them and that are absolutely different. The addition of these graphical objects changes the whole inner world of the applications. (In reality, whatever you see at the screen is a graphical object, but as it was declared in the famous book [1] decades ago: "All animals are equal, but some animals are more equal than others". Controls are those "more equal" elements on the screen and their behaviour is absolutely different from the behaviour of ordinary graphical objects.)

The inheritance of controls from windows is not always obvious, as controls often do not look like windows. Controls have no title bars, so there is no indication that they can be moved; usually there are no borders that indicate the possibility of resizing. But programmers can easily use these features of all the controls and from time to time they do use them, for example, via anchoring and docking. The most important thing is not *how* controls can be moved and resized, but that for them moving and resizing *can be organized*, though I have to mention that the programmers use this moving and resizing of controls as their secret weapon and never give users direct access to it. The designer decides what will be good for users in one or another situation, and, for example, when a user changes the size of the window, then the controls inside can change their size and position, but only according to the decisions previously coded by the designer.

Graphical objects are of an absolutely different origin than controls and, by default, they are neither movable nor resizable. There are ways to make things look different than what they are in reality (programmers are even paid for their knowledge of such tricks). One technique that programmers use is to paint on top of a control: any panel is a control, so it is resizable by default; with the help of anchoring / docking features, it is fairly easy to make an impression as if you have a resizable graphics, which is changing its sizes according to the resizing of the form (dialog). Simply paint on top of a panel and make this panel a subject of anchoring / docking. By default, panels have no visible borders, and if the background color of the panel is the same as its parent form, then there is no way to distinguish between painting in the form or on the panel, which resides on it. Certainly, such "resizing" of graphics is very limited, but in some cases it is just enough; all depends on the

* There are other multi-window operating systems which also use rectangular windows as the basic element. When I write about moving of the windows, it is applied not only to the particular Windows system from Microsoft, but to the whole class of multi-window operating systems. So, in further text, I will not add "*and other similar operating systems*" every time when I mention Windows.

purpose of application. Another solution for resizing of rectangular graphical objects is the use of bitmap operations, but in most cases it cannot be used because of quality problems, especially for enlarging images. Both of these tricky solutions (painting on a panel or using bitmap operations) have one common defect – they can be used only with the rectangular objects.

If any limited area is populated with two different types of tenants (in our case – controls and graphical objects) which prefer to live under different rules, then the only way to organize their peaceful residence and avoid any mess is to force them to live under ONE law. Because graphics are neither movable nor resizable, the easiest solution is to ignore these controls' features, as if they do not exist. That is why so few applications allow users to move around any inner parts. Thus we have **axiom 2**: *On the inner level, objects are usually neither movable nor resizable*. Interestingly, the combined use of these two axioms creates this absolutely paradoxical situation:

- On the upper level, which is not so important for real work, any user has an absolute control of all the components, and any changes are done easily.
- On the inner level, which is much more important for any user because the real tasks are solved here, users have nearly no control at all. If they do have some control, then it is very limited and is always organized indirectly through some additional windows or features.

Axioms I mentioned were never declared as axioms in a strict mathematical way; at the same time I have never seen, read, or heard about even a single attempt to look at this awkward situation any other way than as an axiom and to design any kind of application on a different foundation. Programmers received these undeclared axioms from Microsoft and have worked under these rules for years without questioning them. If you project these same rules on your everyday life, it would be like this: you are free to move around the city or country, but somebody tells you where to put each piece of furniture inside your house. Would you question such a situation?

So, in the world of programs we have a situation when users have to do their most important work inside the applications while being deprived of the real control of these applications; the work can go on only according to the previously developed scenarios. (The questions of whether any form of adaptive interface can really solve the problems are discussed at the beginning of the second part of this book.) I realized years ago that the immovability of all the elements inside the applications became the main problem in design of many programs, but especially in complicated ones. My goal was to find a general solution which would allow to move and resize objects of an arbitrary shape. I had been looking for the general solution and I found it.

## About the structure of this book

There are two main parts in this book:

- Part 1. The design of movable / resizable objects.
- Part 2. The development of user-driven applications.

The first part contains 14 chapters.

Chapter 1 includes the description of the ideas which are transformed into the algorithm for turning any object into movable and resizable.

Chapter 2 describes nodes and covers – the basic level of movability; also the moving of the first real objects – lines – is demonstrated.

Next several chapters either analyse movability of widely used objects of the most popular shapes or describe movements that are often used by objects of different shapes.

Chapter 3 describes various movable rectangles which differ by the type of resizing.

Chapter 4 analyses the basic principles of rotation.

Chapter 5 describes texts in different movements.

Chapter 6 analyses the moving of different polygons (regular, convex, etc.). Also the first example of not the abstract figure, but the real object in moving and rotation is demonstrated here.

Chapter 7 describes the design of the N-node covers which are used in resizing of objects with curved borders.

Chapter 8 analyses movability of objects with the unusual shapes; transparent nodes can be very useful for rings, crescents, sectors, and objects with the holes.

Chapter 9 is about the inner movements – the movements of sliding partitions in already familiar objects - rectangles, circles, and rings. An example of moving an unlimited number of objects of many different shapes is also included into this chapter.

Chapter 10 discusses complex objects, parts of which are involved in individual, related, and synchronous movements. Here questions of identification of objects are discussed. Interesting elements, like track bars, which are widely used in more complex applications are also discussed in this chapter. One example demonstrates a simplified version of scientific applications.

Chapter 11 is about different types of movement restrictions and overlapping prevention. The involvement of many parts and many different objects begins to turn examples into real applications.

All the previous chapters of the first part were devoted to graphical objects; the last three chapters of this part are mostly about controls.

Chapter 12 is about moving / resizing of the solitary controls.

Chapter 13 analyses the widely used pair of elements "control + text" in which the text is represented not as another control (`Label`), but as a painted object.

Chapter 14 is about the movable / resizable groups. Such groups can be organized on absolutely different principles with different levels of fixation for relative positions of elements. The elements of groups can be either pure controls or some combinations of controls and graphical objects, which can influence the design of different groups and produce some unusual solutions.

The discussion of the groups completes the analysis of the movable / resizable elements which can be used in all types of applications. When it is known how to turn into movable / resizable an object of any shape and origin; when it is demonstrated how to organize not only individual, but also the synchronous and related movements of the objects, and after the detailed analysis of movable groups, it is time to turn to the development of applications based on all these elements.

The <u>second part</u> contains five chapters. There are no more small examples to demonstrate one or another detail; this is all about the design of new type of programs, so all the examples are the real applications; some of them are really big.

Chapter 15 postulates the basic rules of user-driven applications and demonstrates their use in the first relatively small application. Another example demonstrates the design on the basis of that class of groups, which I think is the best for development of programs for many different areas. At least, this class of groups is used in all the demonstrated applications beginning from the main form and up to the auxiliary forms at all the levels.

Chapter 16 is about the scientific and engineering programs. The request for the movability of elements was born in this area; this is the area in which I try all my ideas on movability and design; this is the area of my greatest interest throughout my entire professional life. Not surprisingly that the design of user-driven applications in this area is discussed at the most detailed level.

Chapter 17 discusses another attractive area for applying user-driven applications – design of programs for financial and economical analysis. Three types of plots are used in the examples (bar charts, pie charts, and ring sets), but other similar and not very similar plots can be used in the same way. The variety of movable / resizable plots is not the crucial thing; the main discussion is about HOW such applications can be developed on the basis of movable elements.

Chapter 18 is devoted to the transformation of the applications already in use into user-driven ones. How much efforts can be needed? What changes have to be done on the programmers' side? Is it difficult for users to switch from the old style programs to the new one? The example program of this chapter is the one with which everybody is familiar – Calculator.

Chapter 19 demonstrates that the principles of user-driven applications can be very well used even in the areas which are far away from the most serious scientific applications, where they were born.

The book includes the summary of rules for design of movable objects and development of user-driven applications. There are also references and links to other programs and documents, which were developed for better explanation of these items. In addition, there are three appendices.

Appendix A discusses the visualization of covers. Such visualization is never used in the real applications, but it is used in the first part of this book for better explanation of the design of covers.

Appendix B discusses the opportunity of using several movers. There are no examples with more than one mover throughout the entire book (except this appendix), but there are situations when more than one mover is needed.

Appendix C contains the list of all the examples of this book. There are more than 80 different examples; it would be difficult to navigate through and quickly find the needed example without such a reminder. The table includes a small picture of each form and the information about the chapter and the page where its explanation and discussion can be found.

## Preliminary remarks

When any well known object gets new features, the users should be aware of these new things. The goal of my work is to develop an easy to use algorithm for turning ANY screen object into movable and resizable. All those innumerable objects were developed at the best level by their designers, and I do not want in any way to interfere in their design. The appearance of the objects must be exactly the same as it was thought out by their designers. But if there is no visual indication of the new features, then how the users will know that the objects on the screen are movable and resizable? They need to know this fact! Try not to exclaim any astonishment or indignation on reading the previous sentence. I only want to remind you that there is no indication that windows on the upper level can be moved and resized; everyone should know this fact, and this knowledge is enough to navigate throughout the Windows or similar systems. There were no such things before the Windows era, so everyone, who was going to use it for the first time, had to be told beforehand that all those windows could be moved and resized and how he was going to do it. Some people are old enough to remember what was before Windows and at the same time are not old enough to forget it; the younger generation may think that they are already born with the knowledge of movability of windows, but I have a feeling that even they have to be informed about this fact at one moment or another.

After this reminder, I think that the fact that users of the new user-driven applications have to be informed at least once about the movability of each and all objects is not an outrageous one.

Now, if you are aware that all the objects on the screen are movable, then how are you going to move them? The mouse seems to be an obvious and the best instrument, as there are programs in which one or another element is moved, and this is always done with a mouse. Again, the moving of windows is the first example which comes to one's mind. But there is a big difference between the moving of windows on the upper level and my algorithm for all kinds of elements. Each window has a title bar by which it is moved. This decision seems to be good enough when the fact that all the windows have the standard rectangular shape is taken into consideration. I propose the algorithm which works for the objects of an arbitrary shape, so my vision is that objects must be moved by any inner point. And if an object is moved by any point, then there is no need at all in any indication of its movability. You simply should know this fact and that is enough.

If there is no indication of the movability of an object, then how are we going to decide whether an object is movable or not? We do not need to do it: **all the objects must be movable**. I hope that this time you contained your indignation. Have you ever had doubts about the movability of any particular window? So all the objects in the Demo application **WorldOfMoveableObjects** are movable; if any object in this application is not movable, then it was done purposely for better comparison and explanation.

The movability of all the elements is the basis of *user-driven applications*. I have heard the complaints about the underlined statement, but only from the people who never tried the applications on movable / resizable elements. (The famous outcry that "driving the car faster than 15 miles per hour is dangerous for animals and people and thus must be forbidden", that outcry came only from those who never tried the car, but not from those who had ever made a single car trip.)

There are a lot of different objects in our applications; these objects are supposed to be involved in different types of movements; if there are no indications of any movements at all, how are the users going to understand which movements can be applied to each object? This is one of the problems that are mentioned in a lot of examples which I am going to demonstrate and discuss in further chapters.

**Figure I.1** demonstrates the first view that you see on starting the **WorldOfMoveableObjects** application. This is, I think, also the first of my Demo applications in which I did not put on the screen the reminder that ALL the objects in this program (not only in this form, but in all the forms of this application!) are movable. The objects that you see in this figure are discussed in detail in one or another chapter of the book. I think that a quick look across this picture from one object to another may give you a better understanding of what you can find further on. Like a scheme at the entrance of a big museum. Only instead of the "Do not touch the exhibits", you have "Any object can be moved by any inner point", so I do not need to repeat it for each of them.

Polygon in the middle. Configuration of this polygon can be changed by moving any end point of any segment of the perimeter or the central point. By changing the configuration you can literally turn the figure inside out. All other points of the perimeter, except the apices, can be used for scaling the polygon. The polygon can be rotated (right mouse press) by any inner point. Class `ChatoyantPolygon` is discussed in the subsection of the chapter *Polygons*.

| | |
|---|---|
| Group in the top right corner. | The group consists of controls paired with their comments. Any control is moved by the frame; its comment moves synchronously. Comments can be moved independently and placed anywhere. Class `CommentedControl` is discussed in the chapter *Control + text*. A set of elements constitutes the group. The frame of the group adjusts to the positions of all the inner elements and is always shown around them. The title of the group can be moved left and right between the two sides of the group. Class `ElasticGroup` is discussed in the chapter *Groups of elements* and is widely used in all the complex applications demonstrated in the second part of the book. Class `ElasticGroup` has a special tuning form which is used in all other places, but not in this form. |

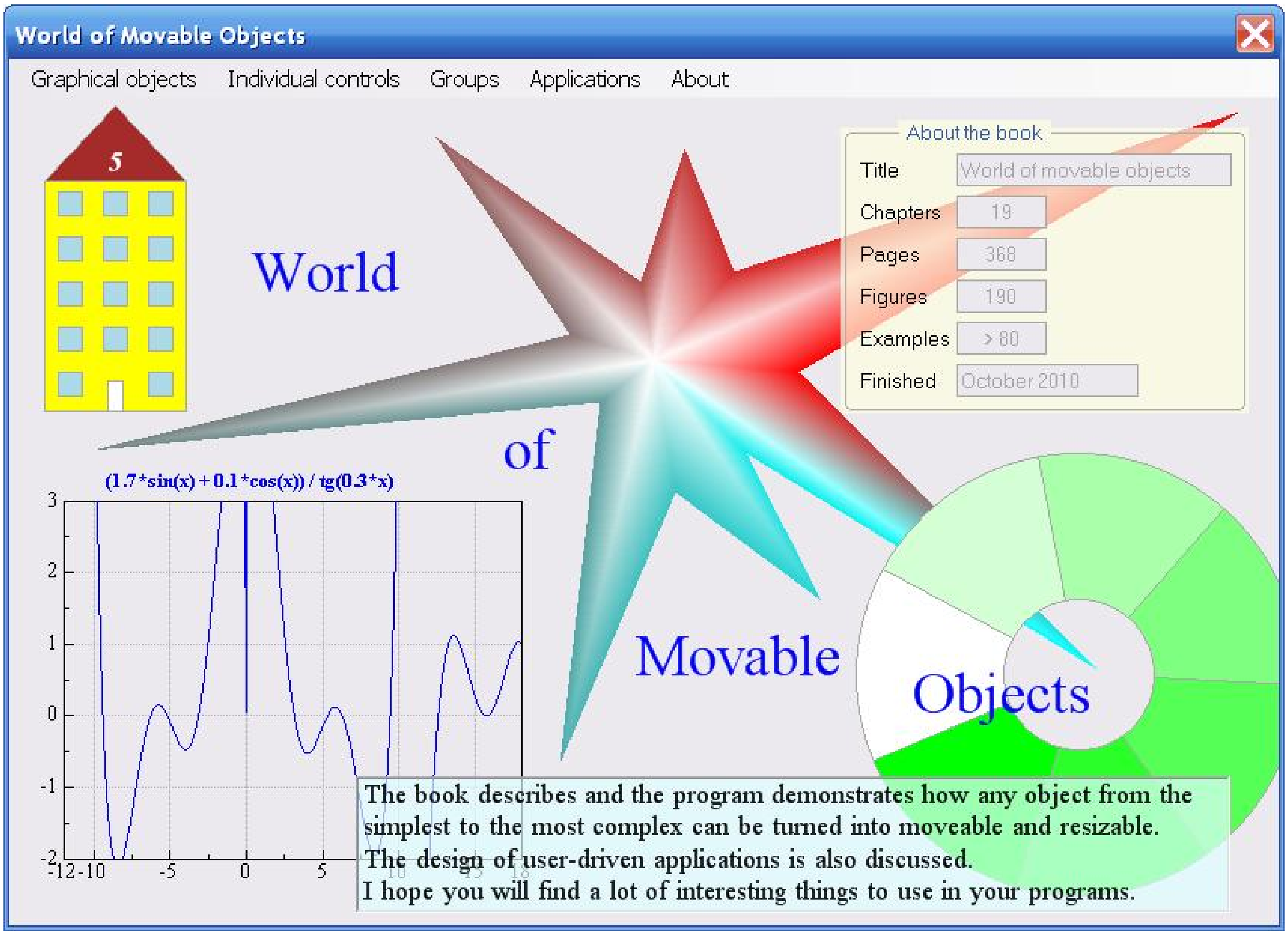

**Fig.I.1** The first view that you will see on starting the application

| | |
|---|---|
| Plot in the bottom left corner. | The main plotting area is resized by borders and corners. The scales can be moved individually and positioned anywhere (with some restrictions, as there must be the conformity between the scales and the main plotting area). Comment belongs to the `CommentToRect` class; the identical copy of this class is discussed in the chapter *Complex objects*. Classes `Scale` and `Plot` are discussed in the chapter *Applications for science and engineering*. Both classes have special tuning forms which are used in all other places (forms), but not in this one. |
| Ring in the bottom right corner. | Ring can be resized by any point of the outer or inner circle; resizing of rings is discussed in the chapter *Curved borders. N-node covers*. The cover of a ring uses a special technique which is discussed in the chapter *Transparent nodes*. This ring has the sliding partitions. Class `PrimitiveRing` is discussed in the chapter *Data visualization*. |
| Information at the bottom. | This text can be moved but not rotated. Class `TextM` is discussed in the chapter *Texts*. |
| Words across the form. | Can be moved and rotated by any point. Class `TextMR` is discussed in the chapter *Texts*. |
| House in the top left corner. | House can be resized by all four sides and all four corners of the rectangular part. The roof top can be moved not only up or down, but also to the sides (no requirement for the symmetry of the roof). Class `SimpleHouse` appears only in the *Appendix B*, but there are similar classes of houses in the chapters *Polygons* and *An exercise in painting*. |

There are two big differences between using objects of the same classes in the **Form_Main.cs** and further on. You can move the objects in the **Form_Main.cs**, you can resize them, but you cannot tune them, and the results are not saved for later use. Design of programs without these two things is against the laws of the user-driven applications, but I decided not to introduce them here. Everything will come at a proper time. Let us start.